\documentclass[letterpaper, 10 pt, conference]{ieeeconf}  % Comment this line out if you need a4paper

\IEEEoverridecommandlockouts                              % This command is only needed if 
\usepackage{amsmath,amssymb,graphicx,chemarr,xcolor,scalerel,mhchem}
\usepackage{comment}
\newtheorem{theorem}{Theorem}

\newtheorem{claim}{Claim}
\renewcommand{\r}{\mathcal{R}}
\renewcommand{\d}{\mathcal{D}}
\newcommand {\B}[1]{\Bar{#1}}
\newcommand{\halmos}{\rule{1ex}{1.4ex}}
\newenvironment{myproof}{\noindent {\em Proof}.\ }{\hspace*{\fill}$\halmos$\medskip}
\newcommand{\epr}{\end{myproof}}
\newcommand{\bpr}{\begin{myproof}}
\title{\LARGE \bf
Multi-variable control to mitigate loads in CRISPRa networks
}

\author{Krishna Manoj$^{1}$, Theodore W. Grunberg$^{2}$, and Domitilla Del Vecchio$^{1}$% <-this % stops a space
\thanks{*This work was supported by NSF CCF-FET Award 2007674}% <-this % stops a space
\thanks{$^{1}$K. Manoj and D. Del Vecchio are at the Department of Mechanical Engineering, MIT, Cambridge, MA 02139, USA.}
\thanks{$^{2}$ T. W. Grunberg is at the Department of Electrical Engineering and Computer Science, MIT, Cambridge, MA 02139, USA.}
\thanks{Emails: {\tt\small kmanoj@mit.edu} (K. Manoj), {\tt\small tgrunber@mit.edu} (T. W. Grunberg),  and {\tt\small ddv@mit.edu} (D. Del Vecchio)}}
\begin{document}

\maketitle
\thispagestyle{empty}
\pagestyle{empty}

%%%%%%%%%%%%%%%%%%%%%%%%%%%%%%%%%%%%%%%%%%%%%%%%%%%%%%%%%%%%%%%%%%%%%%%%%%%%%%%%
\begin{abstract}
The discovery of CRISPR-mediated gene activation (CRISPRa) has transformed the way in which we perform genetic screening, bioproduction and therapeutics through its ability to scale and multiplex. However, the emergence of loads on the key molecular resources constituting CRISPRa by the orthogonal short RNA that guide such resources to gene targets, couple theoretically independent CRISPRa modules. This coupling negates the ability of CRISPRa systems to concurrently regulate multiple genes independent of one another. In this paper, we propose to reduce this coupling by mitigating the loads on the molecular resources that constitute CRISPRa. In particular, we design a multi-variable controller that makes the concentration of these molecular resources robust to variations in the level of the short RNA loads. This work serves as a foundation to design and implement CRISPRa controllers for practical applications. 

\end{abstract}

%%%%%%%%%%%%%%%%%%%%%%%%%%%%%%%%%%%%%%%%%%%%%%%%%%%%%%%%%%%%%%%%%%%%%%%%%%%%%%%%
\section{INTRODUCTION}
Resource competition is a ubiquitous problem in many engineered and natural systems. It occurs when multiple species/entities compete with each other for a shared resource, which is limited in supply. The competition for such shared resources gives rise to loads on the resource itself and thereby to coupling among the entities that compete for the resource. Resource competition has led to major consequences such as the extinction of species \cite{mooney2001evolutionary}, geo-political conflicts between countries \cite{gulley2018china}, and human aggression leading to wars \cite{durham1976resource}. Competition in biomolecular networks takes many forms, from competition for cellular resources by gene expression processes \cite{gyorgy2014limitations,gyorgy2015isocost}, to competition of different transcription factors or proteins for the same binding targets \cite{al2022epigenetic}. In the case of synthetic genetic modules, in particular, these effects may cause performance degradation and loss of modularity.

CRISPR-mediated gene regulation (CRISPRi/a) is one synthetic biology tool that has been shown to be affected by resource competition. CRISPRi/a theoretically allows concurrent regulation of multiple genes through catalytically dead Cas9 (dCas9), which binds to and regulates a target gene after associating with a suitable guide RNA \cite{Bikard2013}. Specific guide RNAs (gRNAs), in turn, are designed specifically for each target gene thereby keeping the regulation of genes independent of each other \cite{larson2013crispr,adli2018crispr}. The toxicity of dCas9, at high concentrations, limits the supply of this molecular resource and thereby induces competition between different gRNA species \cite{Zhang2018}. As a result, and in contrast to what was originally hoped for, current CRISPR/dCas9 systems are not easily scalable nor modular.

During CRISPR-mediated repression (CRISPRi), target-specific gRNAs recruit the limited resource dCas9 to bind to the promoter site of their respective target gene to prevent the binding of RNA polymerase and hence transcription. Recent studies on CRISPRi have investigated the effects of resource competition \cite{chen2018model, anderson2021competitive}, and introduced the design of controllers to mitigate these effects \cite{huang2021dcas9}. Unlike CRISPRi, CRISPR-mediated gene activation (CRISPRa) requires two resources for the activation of a target gene, dCas9, and an RNA binding protein with an activation domain (RBP-AD) \cite{Fontana2018}. The effects of competition for both dCas9 and RBP-AD in CRISPRa networks were characterized in our earlier modeling study \cite{manoj2022emergent}. Specifically, we demonstrated that competition can lead to the degradation of on-target activation performance and to the coupling between theoretically independent modules. In this paper, we address the problem of mitigating competition effects in CRISPRa by designing a biomolecular control architecture that concurrently regulates the free level of dCas9 and RBP-AD to a constant value, robustly to the presence of gRNAs that load these resources.

This paper is organized as follows. We review the effects of competition for two resources in CRISPRa networks in Section \ref{Crispra}. We then focus on the design of a controller for RBP-AD by first assuming no competition for dCas9 in Section \ref{controller}. In Section \ref{dcontroller}, we provide a brief description of the dCas9 controller designed in \cite{huang2021dcas9}. We then, implement both the dCas9 and RBP-AD controllers concurrently and compare the performance of this multi-variable control architecture in Section \ref{ccontroller} to the one of Section \ref{controller}, in which the two resource control problems are artificially decoupled. In Section \ref{sec4}, we conclude the study and provide insights on possible future advancements and developments in the field.

\section{Open-loop behavior of CRISPRa systems \label{Crispra}}
\begin{figure}[ht]
\centering
\includegraphics[width=0.48\textwidth]{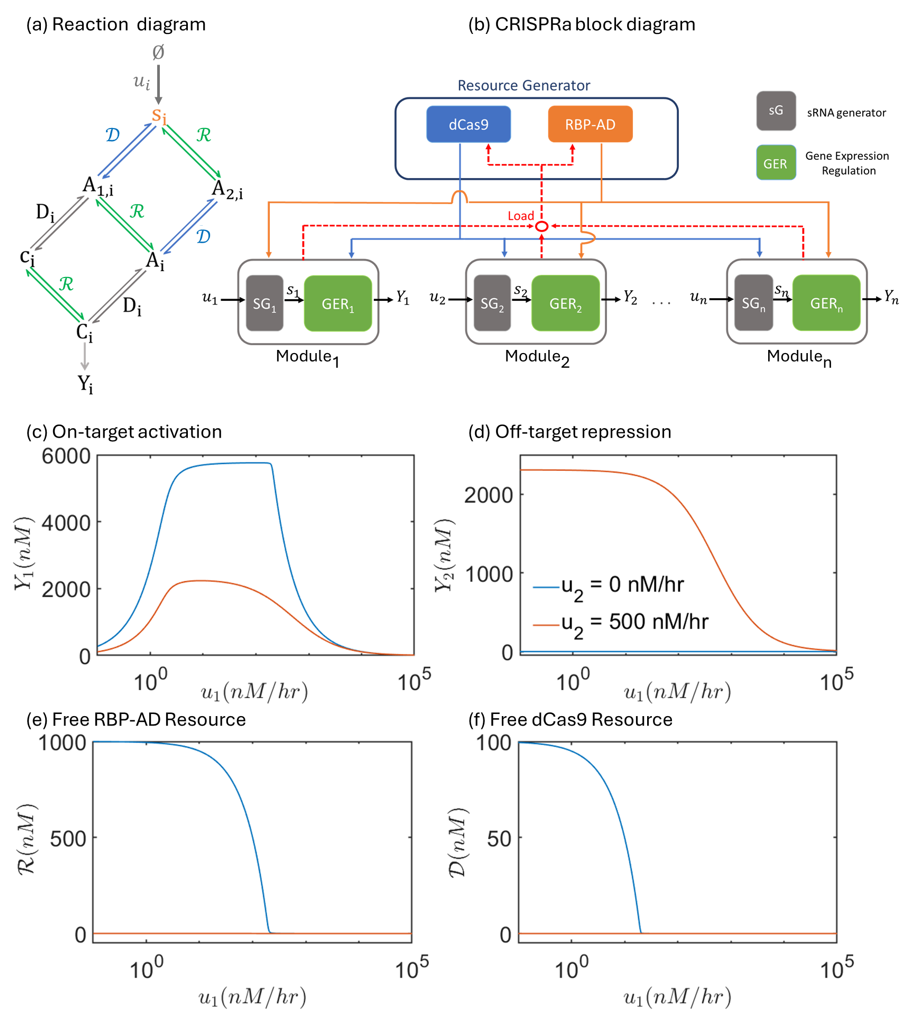}
\caption{CRISPRa network description. (a) Schematic of the double-diamond reaction diagram displaying all the chemical reactions involved for CRISPR-mediated gene activation. (b) Block diagram representation of $n$ parallel CRISPRa modules concurrently utilizing the shared dCas9 and RBP-AD resources along with the loads on the resources due to competition. The concentration levels of (c) output protein $Y_1$ (showing on-target effects), (d) output protein $Y_2$ (showing off-target effects), (e) free RBP-AD resource $\r$ and (f) free dCas9 resource $\d$ as the input $u_1$ is varied with ($u_2 = 500$ nM/hr) and without ($u2 = 0$ nM/hr) competition. The values of all other parameters are given in \cite{manoj2022emergent} and used throughout the paper.} \label{fig:system}
\end{figure}

Here, we review a model that we previously published in \cite{manoj2022emergent}.
The activation of a protein through CRISPRa is a multi-step process involving the guide RNA ($s_i$) that recruits the two protein resources, dCas9 ($\d$) and RBP-AD ($\r$), to activate the target gene $D_i$ for $i \in \{1,n\}$ where $n$ is the total number of CRISPRa modules. The presence of two shared resources gives rise to multiple possible pathways for each binding reaction leading to the activation of the protein. These different pathways can be condensed as shown in Fig. \ref{fig:system} (a) to form a double diamond structure. The guide RNA ($s_i$), produced at a constant rate $u_i$ (input) combines with either dCas9 or RBP-AD to form the intermediate complexes $A_{1,i}$ and $A_{2,i}$, respectively. The intermediate complex $A_{1,i}$ binds with the target gene, $D_i$ (forming $c_i$), and then with RBP-AD to become the transcriptionally active gene $C_i$. The path from $A_{1,i}$ to $C_i$ can also occur by binding with RBP-AD first (forming the CRISPRa complex, $A_i$) and then with the target gene. Moreover, $A_{2,i}$ binds with RBP-AD forming the same CRISPRa complex, $A_i$. The transcriptionally active gene produces the output protein $Y_i$ through transcription and translation.
The reactions describing the system are: 
\begin{align}
&\ce{\mathcal{D} + s_i <=>[\ce{r_i+}][\ce{r_i^-}] \ce{A_{1,i}}} & \quad  \rm \ce{\mathcal{R} + s_i <=>[\ce{p_i+}][\ce{p_i^-}] \ce{A_{2,i}}} \label{rct1}\\
& \ce{A_{1,i} + D_i <=>[\ce{t_i+}][\ce{t_i^-}] \ce{c_i}} & \quad \ce{A_{1,i} + \mathcal{R} <=>[\ce{q_i+}][\ce{q_i^-}] \ce{A_i}} \nonumber\\
&\ce{A_{2,i} + \mathcal{D} <=>[\ce{s_i+}][\ce{s_i^-}] \ce{A_i}} & \quad \ce{c_i + \mathcal{R} <=>[\ce{h_i+}][\ce{h_i^-}] \ce{C_i}} \nonumber \\
&\ce{A_i + D_i <=>[\ce{g_i+}][\ce{g_i^-}] \ce{C_i}} & \quad \ce{C_i ->[\ce{\kappa}] \ce{Y_i + C_i}}. \label{rct2}
\end{align}

We define $\r_t, \d_{t}$ and $D_{it}$ as the total concentrations of RBP-AD, dCas9 and the target gene, respectively, their conservation in the system can be written as:
\begin{align}
    \text{DNA:} & D_{it} = D_i + C_i + c_i \label{start}\\
    \text{dCas9:} & \d_{t} = \d + \Sigma_i C_i + \Sigma_i A_i + \Sigma_i c_i + \Sigma_i A_{1,i} \label{dt}\\
    \text{RBP-AD:} & \r_{t}  = \r + \Sigma_i C_i + \Sigma_i A_i +\Sigma_i  A_{2,i}. \label{rol}
\end{align}

The corresponding reaction rate equations (RREs) can be obtained using mass action kinetics as~\cite{del2014biomolecular}:
\begin{align}
    \Dot{Y_i} & = \kappa C_i - \gamma Y_i \nonumber \\
    \Dot{C_i} & = g_i^+ A_i D_i + h_i^+ c_i \r - (h_i^- +g_i^{-} + \delta) C_i\nonumber \\%\scaleto{-}{3pt}
    \Dot{c_i} & = t_i^+ A_{1,i} D_i  + h_i^- C_i - (t_i^- + h_i^+  \r + \delta)c_i\nonumber \\
    \Dot{A_i} & = q_i^+ A_{1,i} \r + s_i^+ A_{2,i} \d + g_i^- C_i \nonumber\\
    & \qquad- (q_i^- + s_i^- + g_i^+ D_i+ \delta) A_i \nonumber \\
    \Dot{A}_{1,i} & = r_i^+ s_i \d  + q_i^- A_i + t_i^- c_i \nonumber\\
    & \qquad (r_i^- + q_i^+ \r + t_i^+ D_i+ \delta)A_{1,i} \nonumber \\
    \Dot{A}_{2,i} & = p_i^+ \r s_i + s_i^- A_i - (p_i^- + s_i^+ \d+ \delta )A_{2,i} \nonumber \\
    \Dot{s_i} & = u_i + r_i^- A_{1,i} + p_i^- A_{2,i}  - (r_i^+ \d  + p_i^+ \r + \delta) s_i. \label{end}
\end{align}
for $i \in \{1,n\}$, where, $\delta$ and $\gamma$ are the corresponding decay rate constants of gRNAs and proteins.

Figure \ref{fig:system} (b) displays a block diagram with $n$ concurrent CRISPRa modules, each having an input $u_i$ and a corresponding output $Y_i$. Internal to each CRISPRa module, there is a guide RNA generator (denoted as sG) producing $s_i$, which in turn regulates the expression of the protein $Y_i$ (denoted by GER). All CRISPRa modules sequester resources from the shared dCas9 and RBP-AD pool of resources. In the presence of the competitor CRISPRa module, limited resources lead to unwanted coupling between the modules. Specifically, when the input of module$_1$ ($u_1$) increases, the module uses more resources, decreasing the available resource pool for the rest of the modules. This gives rise to an unintended load (shown in red) on the resource-generating unit and, in turn, causes a coupling between otherwise independent modules. 

For $n =2$, Fig. \ref{fig:system} (c) and (d) show the numerical solutions for the steady-state levels of the output proteins $Y_1$ and $Y_2$ as the input $u_1$ varies in a concurrent pair of CRISPRa module. In Fig. \ref{fig:system} (c), we observe a biphasic response in the on-target regulation of the CRISPRa module. The amount of the output protein increases showing the intended activation for low input values, whereas, the amount of the output protein decreases for high input values showing an unintended repression in the system. Such a biphasic response is due to the double diamond structure of the network (Fig. \ref{fig:system}a). The sequestration of the shared resources by the intermediate complexes in the top half of the diamond structure ($A_{1,i}$, $A_{2,i}$ and $c_i$), causes a decrease in the concentration levels of $A_i$, $C_i$ and therefore $Y_i$. Such a behavior is referred to as self-sequestration or squelching \cite{cahill1994regulatory}. Figure \ref{fig:system} (d) shows the decrease in the levels of $Y_2$ as the competitor input $u_1$ is increased for a fixed value of $u_2$. Figure \ref{fig:system} (e) and (f) captures the load on the shared resources, RBP-AD and dCas9, respectively. In the absence of competitors, we observe a monotonic decrease in the resources from their total amounts ($\r_t$ and $\d_t$), showing the rapid sequestration of resources. For high values of $u_2$, both resources are almost completely depleted and remain near zero. 
%\begin{comment}

%\end{comment}

\section{Controller design for RBP-AD} \label{controller}

Having identified the effects of resource competition in the physics-based model in Section \ref{Crispra}, we move on to create a controller to mitigate the effects of competition. Prior research on the competition for dCas9 in CRISPRi networks designed and experimentally implemented a feedback controller of free dCas9 concentration intending to keep free dCas9 level constant irrespective of the load. The result is that any two CRISPRi modules become decoupled from each other despite resource competition \cite{huang2021dcas9}. Our first step here, therefore, is to design a controller for RBP-AD such that the free level of RBP-AD does not significantly change when the inputs to the CRISPRa modules are varied. To this end, we consider two different strategies: buffering and feedback control. Specifically, in Sections \ref{bcontroller} and \ref{fcontroller}, we assume that the concentration of free dCas9 is constant and thereby assume that an ideal dCas9 controller can keep such levels constant independent of the RBP-AD controller. In Section \ref{ccontroller}, we consider the concurrent operation of the dCas9 controller and RBP-AD controller, including the coupling between these two, and evaluate the performance compared to the idealized case considered in this section.

\subsection{Buffering based controller}\label{bcontroller}
\begin{figure}[ht]
\centering
\includegraphics[width=0.48\textwidth]{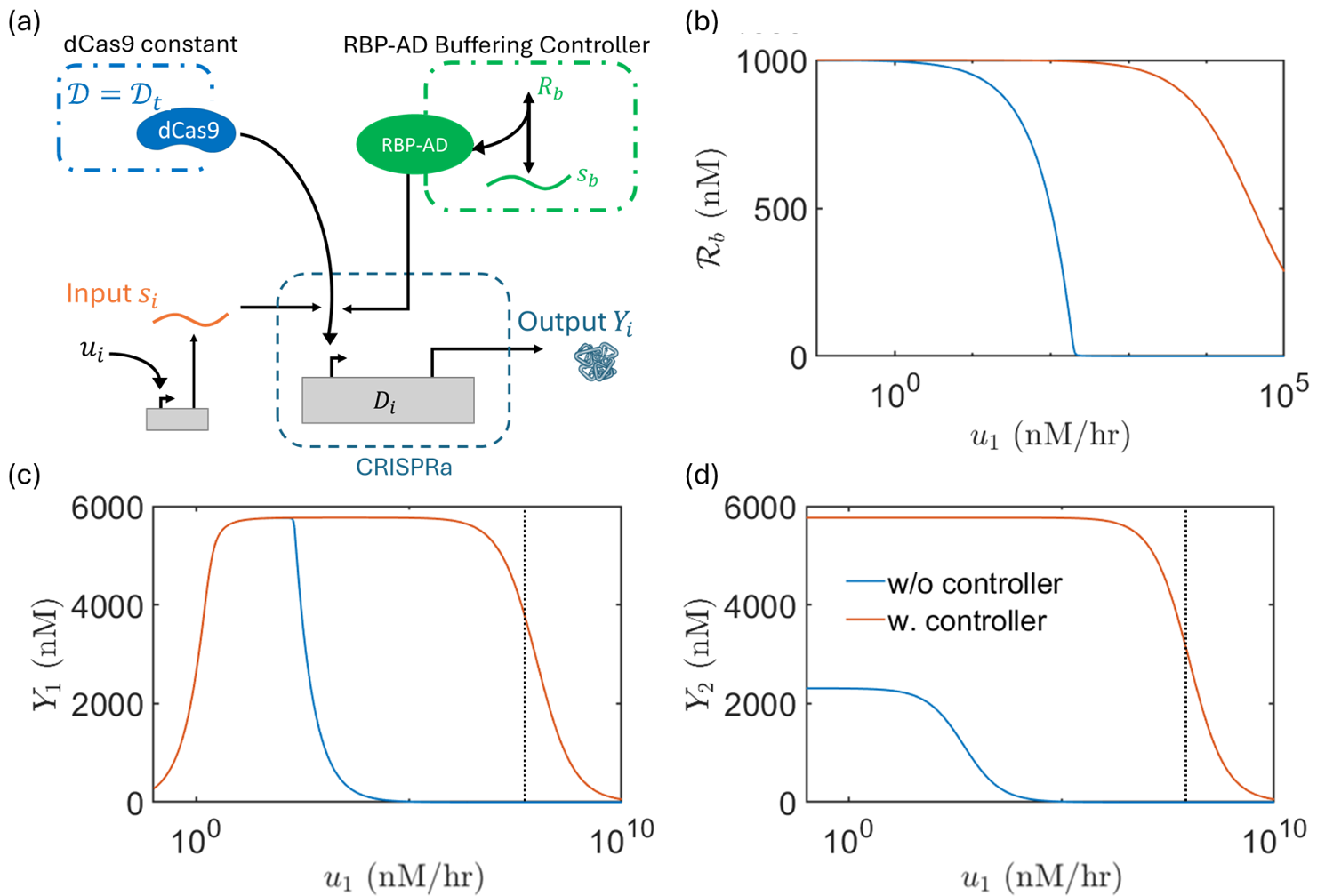}
\caption{Buffering controller attenuates loads for $u_i \ll u_b$ in the absence of dcas9 competition. (a) Genetic circuit diagram for the CRISPRa module with the buffering controller, where the guide RNA $s_b$, sequesters the RBP-AD resource by forming the complex $R_b$. The levels of dCas9 are kept constant and there are $n$ CRISPRa modules functioning in parallel. The concentration levels of (b) free RBP-AD resource $\r_b$ (c) output protein $Y_1$, (d) output protein $Y_2$, as the input $u_1$ is varied with ($u_b = 10^8$ nM/hr) and without ($u_b = 0$ nM/hr) the controller. $u_2 = 0$ nM/hr for (b) and (c), while $u_2 = 500$ nM/hr for (d). The vertical line in (c) and (d) correspond to $u_b = 10^8$ nM/hr.}
\label{fig:buffer}
\end{figure}

The first approach takes advantage of the fact that the resource binding sites for dCas9 and RBP-AD on the guide RNA are disjoint and hence, the binding process of one resource to the guide RNA is not affected by the binding process of the other resource to the same guide RNA. We design an additional guide RNA (denoted as $s_b$), that binds to RBP-AD alone forming the complex $R_b$ (Fig. \ref{fig:buffer}a). The rationale for adding this load through $s_b$ is to create a controllable reservoir of molecules that can be released based on the demand for the resource \cite{hancock2017interplay}. 
To determine the efficacy of the approach, we compare the open-loop system described in Section \ref{Crispra} with free RBP-AD resource denoted as $\r$ to the CRISPRa system with the RBP-AD buffering controller (controlled system), with the free RBP-AD denoted as $\r_b$. The controlled system is still described by the reactions (\ref{rct1})-(\ref{rct2}) in Section \ref{Crispra}, but with $\r$ replaced with $\r_b$, and the additional reactions described below:
\begin{align}
 \it \ce{\r_b + s_b <=>[\ce{p^+}][\ce{p^-}] \ce{R_b}}. \hspace{2ex} 
\end{align}
The RREs are all the ODEs listed in Eq. (\ref{end}) with $\r$ replaced with $\r_b$ as:
\begin{align}
    \Dot{Y_i} & = \kappa C_i - \gamma Y_i \label{bstart}\\
    \Dot{C_i} & = g_i^+ A_i D_i + h_i^+ c_i \r_b - (h_i^- +g_i^{-} + \delta) C_i \label{astart}\\%\scaleto{-}{3pt}
    \Dot{c_i} & = t_i^+ A_{1,i} D_i  + h_i^- C_i - (t_i^- + h_i^+  \r_b + \delta)c_i \nonumber\\
    \Dot{A_i} & = q_i^+ A_{1,i} \r_b + s_i^+ A_{2,i} \d + g_i^- C_i \nonumber\\
    & \qquad - (q_i^- + s_i^- + g_i^+ D_i+ \delta) A_i \nonumber\end{align}\begin{align}
    \Dot{A_{1,i}} & = r_i^+ s_i \d  + q_i^- A_i + t_i^- c_i \nonumber\\
    & \qquad- (r_i^- + q_i^+ \r_b + t_i^+ D_i+ \delta)A_{1,i}  \nonumber\\
    \Dot{A_{2,i}} & = p_i^+ \r_b s_i + s_i^- A_i - (p_i^- + s_i^+ \d+ \delta )A_{2,i} \nonumber\\
    \Dot{s_i} & = u_i + r_i^- A_{1,i} + p_i^- A_{2,i}  - (r_i^+ \d  + p_i^+ \r_b + \delta) s_i \label{aend} \\
    \Dot{R_b} & = p^+ \r_b s_b - (p^- + \delta )R_b \label{R0} \\
    \Dot{s_b} & = u_b + p^- R_b  - (p^+ \r_b + \delta) s_b, \label{sb}
\end{align}
for $i \in \{1,n\}$. The conservation laws for dCas9 and DNA remain unchanged, while, the conservation law for RBP-AD modifies to:
\begin{equation}
    \text{RBP-AD:} \ \r_{total}  = \r_b + \Sigma_i ( C_i + A_i + A_{2,i})+ R_b. \label{rclb}
\end{equation}
where we define: \begin{align}
    \r_{total} = \r_t + \frac{\r_t u_b}{\delta(K_p + \r_t)}, \label{r_tot}
\end{align}
with $K_x = \tfrac{x^- + \delta}{x^+}$ with $x \in \{p,q,r,s,t,g,h\}$ and are the same across CRISPR modules (i.e. $x_i^- = x_j^-$ and $x^+_i = x^+_j$ for all $i\neq j$).
Now we show that the controller with $\r_{total}$ as in (\ref{r_tot}) keeps the free RBP-AD amounts at steady-state, with ($\B{\r}_b$) and without ($\B{\r}$) the controller the same. This is required to make a legitimate comparison between the open-loop and the controlled system. From now on, the steady state value of any concentration level $x$ is denoted as $\B{x}$.
\begin{claim}\label{tm1}
    For the CRISPRa system with the buffering controller governed by (\ref{start}) - (\ref{dt}) and (\ref{bstart}) - (\ref{r_tot}), we have:
    \[ \B{\r}_b = \B{\r} \quad \text{when} \quad u_i = 0 \ \text{for all } i.\]
\end{claim}
\begin{myproof}
    To compare the open-loop and the closed-loop system, we examine the conservation laws for RBP-AD, (\ref{rol}) and (\ref{rclb}). In the absence of any guide RNAs, $u_i = 0,\ \text{for all } i \in \{1,n\}$, the conservation law for the open-loop system, (\ref{rol}), becomes:
\begin{align}
    \B{\r} = \r_t, \label{concwcont}
\end{align}
whereas, the conservation law for the closed-loop system, (\ref{rclb}), becomes:
\begin{align}
    \B{\r}_b + \B{R}_b = \r_{total}. \label{rnewt}
\end{align}
The steady-state levels of $R_b$ is calculated from (\ref{R0}) as:
\begin{align}
    \B{R}_b &= \frac{\B{\r}_b \B{s}_b}{K_p}. \nonumber
    \end{align}
The steady-state levels of total $s_b$, $\B{s}_b + \B{R}_b$ calculated from (\ref{R0}) and (\ref{sb}) is 
    \begin{align}
    \B{R}_b + \B{s}_b &= \B{s}_b \left[ 1 + \frac{\B{\r}_b}{K_p}\right] = \frac{u_b}{\delta} \nonumber\\
    \implies \B{R}_b &= \frac{\B{\r}_b u_b}{\delta(K_p + \B{\r}_b)} \label{R0ss}
\end{align}
%where $K_x = \tfrac{x^- + \delta}{x^+}$ with $x \in \{p,q,r,s,t,g,h\}$.
Substituting (\ref{R0ss}) in (\ref{rnewt}):
\begin{align}
   \B{\r}_b + \frac{\B{\r}_b u_b}{\delta(K_p + \B{\r}_b)}  =  \r_{total} = \r_t + \frac{\r_t u_b}{\delta(K_p + \r_t)}.
\end{align}
As the left-hand side is a strict-monotonic function, there exist a unique solution for the equation, i.e. $\B{\r}_b = \r_t$.
\end{myproof}

Next, we prove that the controller ensures that with high values of $u_b \gg \Sigma_i u_i$, which is the production rate of the controller guide RNA ($s_b$ see (\ref{sb})), corresponding to a high-gain feedback control law,  the controller is able to reduce the competition effects with respect to those in the open loop system. To this end, we determine how $\B{\r}_b$ varies with $\Sigma_i u_i$ as shown in the next theorem.

\begin{theorem} \label{tm2}
    For the CRISPRa system with the buffering controller governed by (\ref{start}) - (\ref{dt}) and (\ref{bstart}) - (\ref{r_tot}), under the assumption of independent binding of RBP-AD (i.e. $K_p = K_q = K_h $), we have:
    \[ \Bar{\r}_b \geq \frac{\r_t}{exp\left(\frac{\Sigma_{i=1}^n u_i}{u_b}\right)},\]
    and the sensitivity of free RBP-AD, $\B{\r}_b$ to total inputs in the system $u_{t} = \Sigma_{i=1}^n u_i$ is given by:
    \[
    \frac{1}{\B{\r}_b}\left|\frac{\mathrm{d} \B{\r}_b}{\mathrm{d} u_t}\right| = \frac{\alpha}{1 + \beta u_t + \frac{u_b}{\delta(K_p + \B{\r}_b)}}\]
where,
\begin{align*}
    \alpha &= \frac{p^+}{\delta(\delta + p^- + p^+ \B{\r}_b)}, \ \text{ and } \  \beta & =  \frac{p^+(\delta + p^-)}{(\delta(\delta + p^- + p^+ \B{\r}_b))^2}.
\end{align*}
\end{theorem}
\begin{myproof}
To obtain the lower bound on $\B{\r}_b$, we define:
\begin{align}
    a_i = s_i + A_{1,i} + c_i \qquad \text{and} \qquad b_i = A_{2,1} + A_i + C_i \label{ab}
\end{align}
Using the new variables, $a_i$ and $b_i$, and assuming independent binding of RBP-AD, (\ref{astart}) - (\ref{aend}) is condensed:
\begin{align}
    \Dot{a_i} & = u_i - p^+ \r_b a_i + p^- b_i - \delta a_i \nonumber\\
    \Dot{b_i} & = p^+ \r_b a_i - p^- b_i -\delta b_i. \label{bdot}
\end{align}
At steady state:
\begin{align*}
    \B{a}_i = \frac{u_i + p^- \B{b}_i}{p^+ \B{\r}_b + \delta} \quad \text{and} \quad \B{b}_i = \frac{p^+ \B{\r}_b u_i}{\delta (\delta + p^- + p^+ \B{\r}_b)}\\
    \implies \Sigma_{i=1}^n \B{b}_i =  \frac{p^+ \B{\r}_b u_t}{\delta (\delta + p^- + p^+ \B{\r}_b)} 
\end{align*}
Substituting in (\ref{rclb}) at steady-state:
\begin{align*}
    &\B{\r}_b + \Sigma_{i=1}^n \B{b}_i + \B{R}_b = \r_{total}\\
    \implies &\r_b + \frac{p^+ \B{\r}_b u_t}{\delta (\delta + p^- + p^+ \B{\r}_b)}  \\
    & \quad + \frac{\B{\r}_b u_b}{\delta(K_p + \B{\r}_b)} - \r_{total} := f_1(\B{\r}_b, u_t, u_b) =0
\end{align*}
Using the implicit function theorem on $f_1(\B{\r}_b, u_t, u_b)$, we get:
\begin{align}
    \frac{\mathrm{d} \B{\r}_b}{\mathrm{d} u_t} &= -\frac{\alpha \B{\r}_b}{1 + \beta u_t + \frac{u_b}{\delta(K_p + \B{\r}_b)}}\geq -\frac{\alpha  \B{\r}_b}{\frac{u_b}{\delta(K_p + \B{\r}_b)}}, \nonumber\\
    \implies \frac{1}{\B{\r}_b}\left|\frac{\mathrm{d} \B{\r}_b}{\mathrm{d} u_t}\right| & = \frac{\alpha}{1 + \beta u_t + \frac{u_b}{\delta(K_p + \B{\r}_b)}} \label{sensb}
\end{align}
The sensitivity tends to zero, as $u_b$ tends to infinity, showing that the variation of $\B{\r}_b$ to the inputs to CRISPRa modules is negligible. Let us define $\r_{min}$ as the solution for:
\[\frac{\mathrm{d} \r_{min}}{\mathrm{d} u_t} = -\frac{\alpha  \B{\r}_b}{\frac{u_b}{\delta(K_p + \B{\r}_b)}} = -\frac{p^+ \r_{min} (K_p + \r_{min})}{u_b(\delta + p^- + p^+ \r_{min})}\]
Hence, we have that:
\begin{align}
    \frac{\mathrm{d} \B{\r}_b}{\mathrm{d} u_t} \geq \frac{\mathrm{d} \r_{min}}{\mathrm{d} u_t}, \label{der}
\end{align}
then $\r_{min}$ is the minimum amount of $\B{\r}_b$ for all $u_i$, $\B{\r}_b\geq \r_{min}$.
The above inequality can be proved by contradiction, assuming there exists a $u_i$ for which $\B{\r}_b < \r_{min}$ satisfies (\ref{der}). $\B{\r}_b < \r_{min}$ implies the trajectories of $\B{\r}_b$ and $\r_{min}$ with $u_i$ would cross at a point. Computing the derivative $\tfrac{\mathrm{d} \B{\r}_b}{\mathrm{d} u_t}$ at the point of the crossing would not satisfy (\ref{der}), which is a contradiction.

Integrating the above ODE for $\r_{min}$ with the limits of 0 to $u_t$ and $\r_t$ to $\r_{min}$, we get:
\begin{align*}
    \frac{\r_t}{\r_{min}} & = exp\left(\frac{u_t}{u_b}\right)
    \implies \r_{min} &= \frac{\r_t}{exp\left(\frac{u_t}{u_b}\right)}
\end{align*}
Under the limit of $u_b \rightarrow \infty$: 
\[\lim_{u_b \to \infty} \r_{min} = \lim_{u_b \to \infty} \frac{\r_t}{exp\left(\frac{u_t}{u_b}\right)} = \r_t. \]
\end{myproof}

This theorem states that as we increase the feedback ``gain'' $u_b$, the free RBP-AD level approaches $\r_t$, thereby attenuating the effect of the loads $u_i$ on the free RBP-AD concentration. 

Figure \ref{fig:buffer} (b) shows the variation of the steady-state levels of free resource RBP-AD ($\B{\r}_b$) as $u_1$ is varied for the case of $n=1$ assuming $\d = \d_t$. The concentration levels of free RBP-AD resource are regulated such that the competition effects are mitigated for $u_b \gg u_1$. For the on-target activation in the absence of a competitor (Fig. \ref{fig:buffer}c), self-sequestration is also evaded when $u_b \gg u_1$, and commences as $u_1$ approaches $u_b$. For $n=2$, Fig. \ref{fig:buffer} (d) shows the off-target response for a constant $u_2$. For $u_b \gg u_1$, the variation in $Y_2$ with $u_1$ is negligible. Note: The simulations are performed for $u_b = 10^8$ nM/hr to ensure that $u_b \gg u_i$. The value of $u_b$ may be altered based on specific experimental properties (maximum transcription rate) without hindering the controller activity as long as $u_b \gg u_i$ is satisfied.

\subsection{Feedback controller} \label{fcontroller}

\begin{figure}[ht]
\centering
\includegraphics[width=0.495\textwidth]{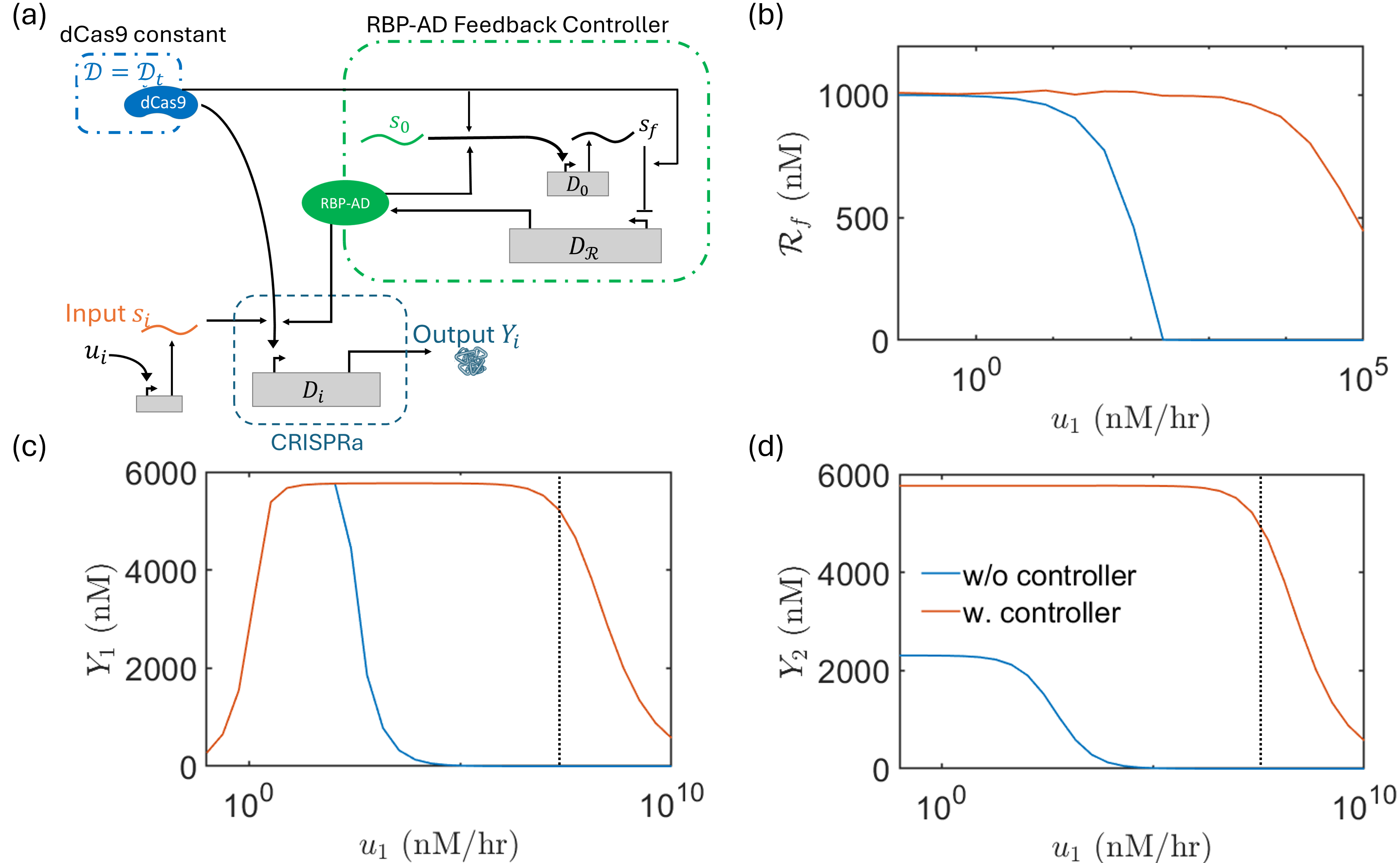}
\caption{Feedback controller performs better than buffering controller in the absence of dcas9 competition. (a) Genetic circuit diagram for the CRISPRa module with the feedback controller, where the guide RNA $s_0$, activates the production of the repressing guide RNA $s_f$ that regulates the production of RBP-AD. The levels of dCas9 are kept constant and there are $n$ CRISPRa units functioning in parallel. The concentration levels of (b) free RBP-AD resource $\r_f$ (c) output protein $Y_1$, (d) output protein $Y_2$, as the input $u_1$ is varied with ($u_0 = 10^8$ nM/hr) and without ($u_0 = 0$ nM/hr) the controller. $u_2 = 0$ nM/hr for (b) and (c), while $u_2 = 500$ nM/hr for (d). The vertical lines in (c) and (d) correspond to $u_0 = 10^8$ nM/hr.}
\label{fig:feedback}
\end{figure}

The RBP-AD controller consists of a CRISPRa unit which activates a CRISPRi module that negatively regulates the production of RBP-AD (Fig. \ref{fig:feedback}a). In particular, the CRISPRa unit has guide RNA $s_0$ activating the gene $D_0$ that expresses the guide RNA $s_f$, which recruits dCas9 to repress the production of RBP-AD from the RBP-AD gene $D_\r$. In this manner, a feedback loop is introduced to regulate the concentration of RBP-AD. The CRISPRa system with the feedback controller is referred to as the closed-loop system. We denote the amount of free RBP-AD resource present in this system as $\r_f$, where subscript \lq$f$\rq~stands for feedback controller. The reactions involved are those listed in (\ref{rct1})-(\ref{rct2}) with $\r$ replaced with $\r_f$ and the following ones.
\begin{align}
&\ce{\mathcal{D} + s_0 <=>[\ce{r_0+}][\ce{r_0^-}] \ce{A_{1,0}}} & \quad  \rm \ce{\mathcal{R}_f + s_0 <=>[\ce{p_0+}][\ce{p_0^-}] \ce{A_{2,0}}} \nonumber\\
& \ce{A_{1,0} + D_0 <=>[\ce{t_0+}][\ce{t_0^-}] \ce{c_0}} & \quad \ce{A_{1,0} + \mathcal{R}_f <=>[\ce{q_0+}][\ce{q_0^-}] \ce{A_0}} \nonumber\\
&\ce{A_{2,0} + \mathcal{D} <=>[\ce{s_0+}][\ce{s_0^-}] \ce{A_0}} & \quad \ce{c_0 + \mathcal{R}_f <=>[\ce{h_0+}][\ce{h_0^-}] \ce{C_0}} \nonumber \\
&\ce{A_0 + D_0 <=>[\ce{g_0+}][\ce{g_0^-}] \ce{C_0}} & \quad \ce{C_0 ->[\ce{\kappa}] \ce{s_f + C_0}}. \nonumber \\
&\ce{\mathcal{D} + s_f <=>[\ce{r_f+}][\ce{r_f^-}] \ce{A_{1,f}}} & \quad  \rm \ce{A_{1,f} + D_{\mathcal{R}} <=>[\ce{t_f+}][\ce{t_f^-}] \ce{c_f}} \nonumber\\
&\ce{D_{\mathcal{R}} ->[\ce{\kappa_r}] \ce{\mathcal{R}_f + D_{\mathcal{R}}}}. &
\end{align}

The corresponding RREs for the CRISPRa system with a feedback controller are:
\begin{align}
    \Dot{Y_j} & = \kappa C_j - \gamma Y_j \label{fstart}\\
    \Dot{C_i} & = g_i^+ A_i D_i + h_i^+ c_i \r_f - (h_i^- +g_i^{-} + \delta) C_i \nonumber\\%\scaleto{-}{3pt}
    \Dot{c_i} & = t_i^+ A_{1,i} D_i  + h_i^- C_i - (t_i^- + h_i^+  \r_f + \delta)c_i \nonumber\\
    \Dot{A_i} & = q_i^+ A_{1,i} \r_f + s_i^+ A_{2,i} \d + g_i^- C_i \nonumber\\
    & \qquad - (q_i^- + s_i^- - g_i^+ D_i+ \delta) A_i \nonumber\\
    \Dot{A_{1,i}} & = r_i^+ s_i \d  + q_i^- A_i + t_i^- c_i \nonumber\\
    & \qquad- (r_i^- + q_i^+ \r_f + t_i^+ D_i+ \delta)A_{1,i}  \nonumber\\
    \Dot{A_{2,i}} & = p_i^+ \r_f s_i + s_i^- A_i - (p_i^- + s_i^+ \d+ \delta )A_{2,i} \nonumber\\
    \Dot{s_i} & = u_i + r_i^- A_{1,i} + p_i^- A_{2,i}  - (r_i^+ \d  + p_i^+ \r_f + \delta) s_i \nonumber \\
    \Dot{s_f} & = \kappa_0 C_0 + r_f^- A_{1,f} - (r_f^+ \d + \delta) s_f \nonumber\\
    \Dot{A_{1,f}} & = r_f^+ s_f \d  + t_f^- c_f - (r_f^- + t_f^+ D_\r+ \delta)A_{1,f} \nonumber \\
    \Dot{c_f} & = t_f^+ A_{1,f} D_\r  - (t_f^- + \delta)c_f \nonumber\\
    \Dot{\r_f} & = \kappa_r D_\r + \Sigma_{i=0}^n (p_i^- A_{2,i} + q_i^- A_i + h_i^- C_i) \nonumber\\
    &\qquad -  \Sigma_{i=0}^n(p_i^+ s_i + q_i^+ A_{1,i} + h_i^+ c_i)\r_f - \gamma \r_f \nonumber
\end{align}
for $i \in \{0,1,2,...,n\}$ and $j \in \{1,n\}$.
The conservation laws for the system are:
\begin{align*}
    &\text{DNA: }  D_{it} = D_i + C_i + c_i\\
    &\text{RBP-AD gene: }   D_{\r t} = D_\r + c_f\\
    &\text{dCas9: }  \d_{t} = \d + \Sigma_i (C_i + A_i +  c_i + A_{1,i} ) + A_{1,f} + c_f.
\end{align*}
We set the production rate of $\r_f$ as: \begin{align}
    \kappa_r &= \left[\r_t + \frac{\r_t u_0}{\delta(K_p + \r_t)}\right] \frac{\gamma(1 + \frac{{\d_t \mathcal{Y} }}{K_r K_t})}{D_{\r t}} \label{rf_tot}.
\end{align}
where $\mathcal{Y}= \mathcal{Y}(\r_t)$ is the function defined as the steady-state concentration of the output from an isolated CRISPRa module with input $u_0$, and with free RBP-AD and free dcas9 held constant at $\r_t$ and $\d_t$, respectively.  

Now, we show that with $\kappa_r$ as in (\ref{rf_tot}) keeps the RBP-AD concentration of the closed-loop system the same as that of the open-loop system when CRISPRa modules are absent. This is done to ensure the two nonlinear systems are comparable when operating at about the same nominal point.
\begin{claim}\label{tm3}
    For the CRISPRa system with the feedback controller governed by (\ref{fstart}) -  (\ref{rf_tot}), we have:
    \[ \B{\r}_f = \r_t \quad \text{when} \quad u_i = 0 \ \text{for all } i \in \{1,n\}.\]
\end{claim}
\begin{myproof}
    To compare the open-loop and the closed-loop system, we examine the conservation laws for RBP-AD, (\ref{rol}) with that of the closed-loop at steady state. In the absence of any guide RNAs, $u_i = 0,\ \text{for all } i\in \{1,n\}$, the conservation law for the open-loop system, (\ref{rol}), is $\B{\r} = \r_t$. The conservation law for the closed-loop system at steady state is:
\begin{align*}
    \B{\r}_f + \Sigma_{i=0}^n \left[\B{A}_{2,i} + \B{A}_i + \B{c}_i \right] = \frac{\kappa_r \B{D}_\r}{\gamma}.
\end{align*}
For $u_i = 0$ for $i \in \{1,n\}$, the conservation law becomes:
\begin{align*}
    \B{\r}_f + \frac{\B{\r}_f u_0}{\delta (K_p + \B{\r}_f)} = \frac{\kappa_r \B{D}_\r}{\gamma} = \frac{\kappa_r \B{D}_{\r t}}{\gamma (1 + \overline{A}_{1,f})}\\
    \implies \kappa_r = \left[\B{\r}_f + \frac{\B{\r}_f u_0}{\delta(K_p + \B{\r}_f)}\right] \frac{\gamma(1 + \frac{\overline{A}_{1,f}}{K_t})}{D_{\r t}} 
\end{align*}

The right-hand side being a strict-monotonic function implies $\B{\r}_f = \r_t$.
\end{myproof}

Now, we characterize the performance of this controller in regulating $\r_f$.
\begin{theorem} \label{tm3}
    For the CRISPRa system with the feedback controller governed by (\ref{fstart}) -  (\ref{rf_tot}), under the assumption of independent binding of RBP-AD ($K_p = K_q = K_h $), we have:
    \[ \Bar{\r}_f \geq \frac{\r_t K_p}{(\r_t + K_p)exp\left(\frac{u_t}{u_0}\right) - \r_t}\]
    %where $r_{fmin}$ satisfies:
    %\[\frac{r_t}{r_{fmin}}\frac{r_{fmin}  + K_p}{r_t + K_p}  = exp\left(\frac{\Sigma_{i=1}^n u_i}{u_0}\right),\]
    %\[r_{fmin} = \frac{r_t K_p}{(r_t + K_p)exp\left(\frac{\Sigma_{i=1}^n u_i}{u_0}\right) - r_t},\]
    and the sensitivity of free RBP-AD, $\B{\r}_f$ to $u_t$ is given by:
    \[\frac{1}{\B{\r}_f}\left|\frac{\mathrm{d} \B{\r}_f}{\mathrm{d} u_t}\right| = \frac{\alpha_f}{1 + \beta_f u_t  + \beta_f u_0 + \Delta_f} \]
where
\begin{align*}
     \alpha_f &= \frac{1}{\delta (K_p + \B{\r}_f)}, \qquad \beta_f  = \frac{K_p}{\delta (K_p + \B{\r}_f)^2},\\
    \text{and }\Delta_f & = \frac{\kappa_r D_{\r t} \B{\d}}{\gamma K_r ( 1 + \frac{\B{A}_{1,f}}{K_t})^2} \left. \frac{\mathrm{d} \mathcal{Y}}{\mathrm{d} \r_t}\right|_{\r_f} > 0.
\end{align*}
\end{theorem}
\begin{myproof}
    From the RRE of $\r_f$ and conservation law for $D_\r$:    
    \begin{align*}
        \gamma \B{\r}_{f} +& p^+  \B{\r}_{f}\Sigma_{i=0}^n (s_i + A_{1,i} + c_i) \\
        & \qquad - p^- \Sigma_{i=0}^n (A_{2,i} + A_{i} + C_i) = \kappa_r \B{D}_\r
    \end{align*}
Let $a_i$ and $b_i$ be defined as shown in (\ref{ab}) and performing a similar approach as theorem \ref{tm2},
\[\gamma \B{\r}_{f} + p^+  \B{\r}_{f} \Sigma_{i=0}^n a_i - p^- \Sigma_{i=0}^n b_i = \kappa_r \B{D}_\r. \]
The steady-state equation for $b_i$ (similar to (\ref{bdot})) gives $p^+  \B{\r}_{f} a_i - p^- b_i = -\delta b_i$. Substituting $b_i$, we obtain:
    \begin{align}
       \gamma \B{\r}_f + \frac{\B{\r}_f u_t}{K_p + \B{\r}_f} &= \kappa_r D_\r = \kappa_r \frac{D_{\r t} K_t}{K_t + \B{A}_{1,f}} \nonumber\\
        \B{A}_{1,f} & = \frac{\B{s}_f \B{\d}}{K_r} = \frac{\B{\d}}{K_r} \left.\mathcal{Y}(\r_t)\right|_{\B{\r}_f}. \label{imp2}
    \end{align}
Applying the implicit function theorem on (\ref{imp2}), we obtain:
\begin{align}
    \frac{\mathrm{d} \B{\r}_f}{\mathrm{d} u_t} = -\frac{\alpha_f \B{\r}_f}{1 + \beta_f (u_t + u_0) + \Delta_f \frac{\mathrm{d} \mathcal{Y}}{\mathrm{d} \B{\r}_f}} \geq -\frac{\alpha}{\beta_f u_0}. \label{senrf}\\
    \implies \frac{1}{\B{\r}_f}\left|\frac{\mathrm{d} \B{\r}_f}{\mathrm{d} u_t}\right| = \frac{\alpha_f}{1 + \beta_f (u_t + u_0) + \Delta_f \frac{\mathrm{d} \mathcal{Y}}{\mathrm{d} \B{\r}_f}}. \label{senf}
\end{align}
Same as before, the derivative of $\frac{\mathrm{d} \B{\r}_f}{\mathrm{d} u_t}$ tends to zero, as $u_0$ tends to infinity, showing that the sensitivity of $\B{\r}_f$ to the inputs to CRISPRa modules is negligible. Let $\r_{fmin}$ be defined as the solution for:
\[\frac{\mathrm{d} \r_{fmin}}{\mathrm{d} u_t} = -\frac{\frac{\B{\r}_f}{\delta (K_p + \B{\r}_f)}}{\frac{K_p}{\delta (K_p + \B{\r}_f)^2} u_0} = -\frac{\B{\r}_f(K_p + \B{\r}_f)}{K_p u_0}\]
Using a similar approach as before, 
\begin{align*}
    \frac{\mathrm{d} \B{\r}_f}{\mathrm{d} u_t} \geq \frac{\mathrm{d} \r_{fmin}}{\mathrm{d} u_t}, 
\end{align*}
implies that $\B{\r}_f\geq \r_{fmin}$.
Integrating the above ODE for $\r_{fmin}$ with the limits of 0 to $u_t$ and $\r_t$ to $\r_{fmin}$, we get:
    \begin{align*}
         \frac{\r_{t} }{\r_{fmin}} \frac{\r_{fmin}  + K_p}{\r_t + K_p}  &= exp\left(\frac{u_t}{u_0}\right),\\
    \implies \r_{fmin} &= \frac{\r_t K_p}{(\r_t + K_p)exp\left(\frac{u_t}{u_0}\right) - \r_t}.
    \end{align*}
Under the limit of $u_0 \to \infty$:
\[\lim_{u_0 \to \infty} \r_{fmin} = \lim_{u_0 \to \infty} \frac{\r_t K_p}{(\r_t + K_p)exp\left(\frac{u_t}{u_0}\right) - \r_t} = \r_t.\]
\end{myproof}

This theorem states that as we increase the feedback gain $u_0$, the free RBP-AD concentration approaches a uniform value independent of the loads $u_i$, just like in the buffering case (Theorem \ref{tm2}). Figure \ref{fig:feedback} (b) shows the variation of the steady-state levels of free resource RBP-AD ($\B{\r}_f$) as $u_1$ is varied for the case of $n=1$ for $\d = \d_t$. We observe that the feedback controller regulates $\B{\r}_f$ and the resources start to deplete at a larger $u_i$. In Fig. \ref{fig:feedback} (c), self-sequestration is observed as $u_i$ approaches $u_0$. For the case $n=2$, Fig. \ref{fig:feedback} (d) shows negligible off-target effects as the output $Y_2$ remains constant irrespective of $u_1$ for a constant $u_2$ and $u_0\gg u_i$. 

\begin{figure}[ht]
\centering
\includegraphics[width=0.49\textwidth]{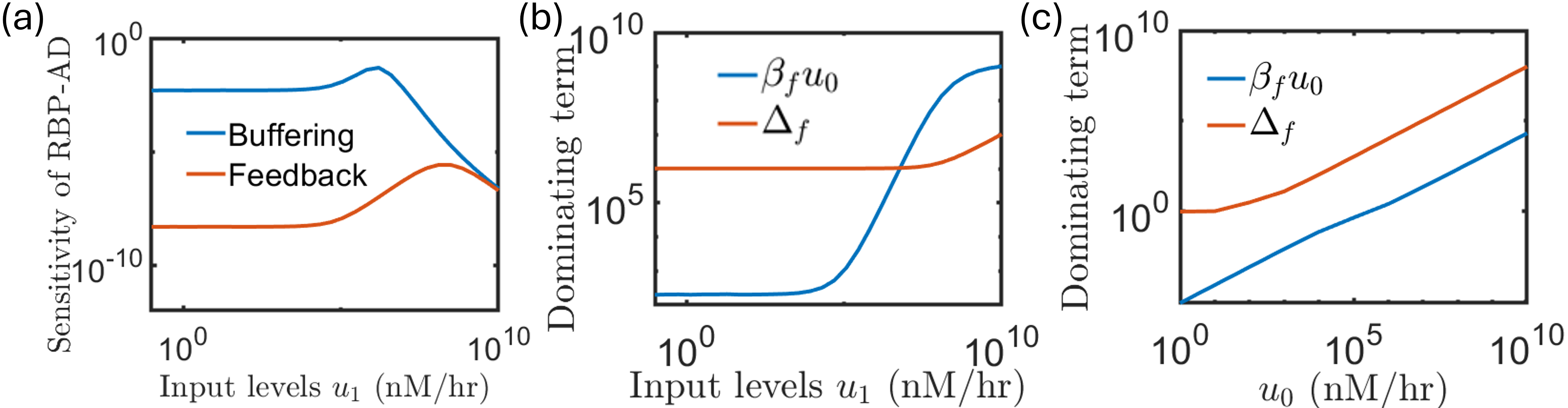}
\caption{Sensitivity comparison between buffering and feedback controllers for RBP-AD. (a) Sensitivity of free RBP-AD using the buffering controller ($\r_b$ from (\ref{sensb})) and feedback controller ($\r_f$ from (\ref{senf})) as $u_1$ is varied for $u_0 = u_b = 10^{10}$ nm/hr. Variation of the terms $\beta_f u_0$ corresponding to the buffering controller (and the buffering section of the feedback controller) and $\Delta_f$ corresponding to the feedback loop in the feedback controller with (b) loads $u_1$ and (c) controller gain $u_0$.}
\label{sensitivity}
\end{figure}

The sensitivity of the feedback controller (\ref{senf}) is decreased due to the added term $\Delta_f$ in the denominator when compared to the buffering controller. This change in sensitivity is illustrated in Fig. \ref{sensitivity}. The sensitivity of $\B{\r}_f$ to $u_i$ in the feedback controller is lower than the sensitivity of $\B{\r}_b$ to $u_i$ in the buffering controller for low values of the inputs, $u_i$ (Fig. \ref{sensitivity}a). As the input $u_i$ is increased, the sensitivities of both controllers coincide.  This is further verified in Fig. \ref{sensitivity} (b), where the terms in the sensitivity equation responsible for the feedback control contribution is $\Delta_f$ and the buffering contribution is $\beta_f u_0$. For small $u_i$, the feedback contribution of the controller dominates the regulation. On the other hand, the buffering contribution in the controller dominates for high $u_i$. Furthermore, analyzing the behavior of the sensitivity term in the feedback controller for increasing controller gain $u_0$, the feedback term dominates the regulation dynamics irrespective of $u_0$ (Fig. \ref{sensitivity}c).

We conclude that the feedback controller outperforms the buffering controller when competition for dCas9 is nonexistent, i.e., dCas9 is maintained constant. So far, we have established that both the buffering controller and the feedback controller are effective in making the concentration levels of RBP-AD robust to loads under constant dCas9 availability. In the next section, we introduce the dCas9 controller published in \cite{huang2021dcas9} and combine it with the RBP-AD controller designed here.

\begin{figure}[ht]
\centering
\includegraphics[width=0.47\textwidth]{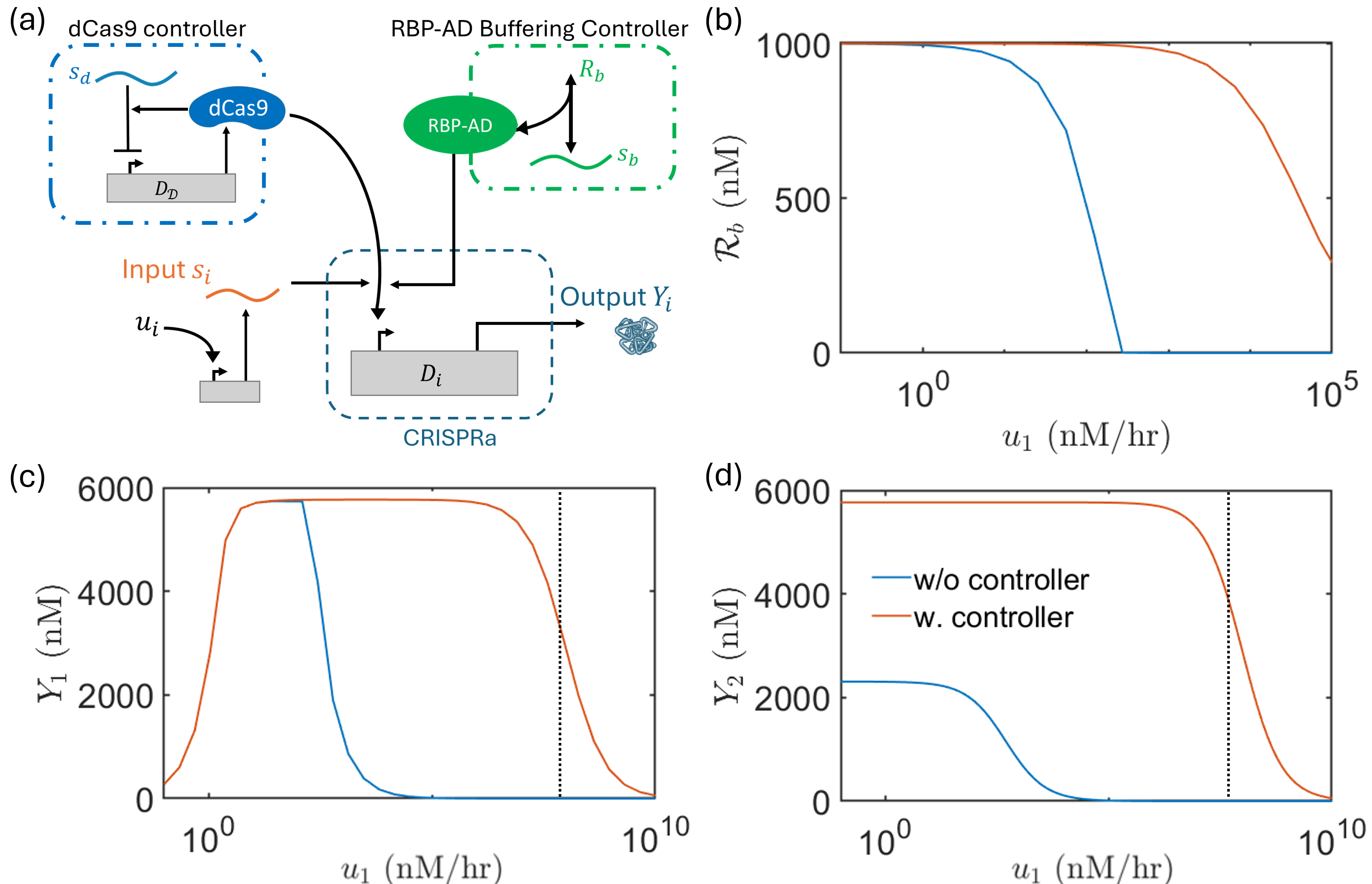}
\caption{Efficacy of the buffering controller for RBP-AD with the dCas9 controller designed in \cite{huang2021dcas9}. (a) Genetic circuit diagram for the CRISPRa module with the buffering controller for RBP-AD (from Section \ref{bcontroller}) and feedback controller for dCas9 (from Section \ref{dcontroller}).The concentration levels of (b) free RBP-AD resource $\r_b$ (c) output protein $Y_1$, (d) output protein $Y_2$, as the input $u_1$ is varied with ($u_b = 10^8$ nM/hr and $u_d = 10^{10}$ nM/hr) and without ($u_b = 0$ nM/hr and $u_d = 0$ nM/hr) the controller. $u_2 = 0$ nM/hr for (b) and (c), while $u_2 = 500$ nM/hr for (d). The vertical line in (c) and (d) correspond to $u_b = 10^8$ nM/hr.}
\label{fig:buffer:dcas9}
\end{figure}

\section{Combined $\rm d$C$\rm as$9 and RBP-AD controller}
\subsection{Brief review of the $\rm d$C$\rm as$9 controller} \label{dcontroller}
Here, we review the design of a controller to regulate free dCas9 levels in the face of loads designed in \cite{huang2021dcas9}. The design of the dCas9 controller involves an additional guide RNA, $s_d$ that recruits dCas9 ($\d_f$, where subscript \lq$f$\rq~stands for feedback control) and targets the promoter for dCas9 ($D_\d$). In this manner, at high dCas9 levels, $s_d$ represses the dCas9 production thereby creating a negative feedback loop. The reactions involved are:
\begin{align*}
&\ce{\mathcal{D}_f + s_d <=>[\ce{r_d+}][\ce{r_d^-}] \ce{A_{1,d}}} & \quad  \ce{A_{1,d} + D_{\mathcal{D}} <=>[\ce{t_d+}][\ce{t_d^-}] \ce{c_d}}\\
&\ce{c_d ->[\ce{r_d^-}] \ce{\mathcal{D}_f + s_d + D_{\mathcal{D}}}} & \quad \ce{D_{\mathcal{D}} ->[\ce{\kappa_d}] \ce{\mathcal{D}_f + D_{\mathcal{D}}}}. 
\end{align*}
and the corresponding RREs are:
\begin{align*}
    \Dot{s_d} & = u_d + r_d^- A_{1,d} + r_d^- c_d - (r_d^+ \d_f + \delta) s_d \nonumber\\
    \Dot{A_{1,d}} & = r_d^+ s_d \d_f  + t_d^- c_d - (r_d^- + t_d^+ D_\d+ \delta)A_{1,d} \nonumber \\
    \Dot{c_d} & = t_d^+ A_{1,d} D_\d  - (t_d^- + r_d^- + \delta)c_d \nonumber \\
     \Dot{\d_f} & = \kappa_d D_\d + r_d^- A_{1,d} - (r_d^+ +  \gamma)\d_f \nonumber 
\end{align*}
where:
\begin{align*}
    \kappa_d &= \left(\d_t + \B{A}_{1,d} + \frac{\B{A}_{1,d} D_{\d t}}{(K_t + \B{A}_{1,d})}\right) \frac{\gamma (K_t + \B{A}_{1,d})}{D_{\d t}K_t}
\end{align*}
with conservation law, $D_{\d t} = D_\d + c_d$.

\subsection{Concurrent operation of $\rm d$C$\rm as$9 and RBP-AD controllers}\label{ccontroller}

In this section, we examine the behavior of the buffering and feedback controller designed in Section \ref{controller} with the additional dCas9 controller introduced in Section  \ref{dcontroller}. Although the dCas9 controller attenuates the loads on the resource, it does not completely remove resource competition from the system. Therefore, the coupling of the two resources may affect the performance of each of these controllers. Specifically, as the feedback controller designed in Section \ref{fcontroller} uses dCas9 for both its CRISPRa and CRISPRi modules, the presence of dCas9 competition is unfavorable. Hence, the performance of the buffering and the feedback controllers for RBP-AD with the dCas9 controller requires examination.

The buffering controller, given in Section \ref{bcontroller} is by design independent of the levels of dCas9 (Fig. \ref{fig:buffer:dcas9}a). Therefore Claim 1 and Theorem 1 apply, even with the addition of the dCas9 controller. The RBP-AD concentration in Figure \ref{fig:buffer:dcas9} (b) with the dCas9 controller compared with Fig. \ref{fig:buffer} (b) with constant dCas9 show negligible change. Further, the behavior of output protein $Y_1$ (compare Fig. \ref{fig:buffer}c and Fig. \ref{fig:buffer:dcas9}c) and $Y_2$ (compare Fig. \ref{fig:buffer}d and Fig. \ref{fig:buffer:dcas9}d) also show similar behavior.

\begin{figure}[ht]
\centering
\includegraphics[width=0.49\textwidth]{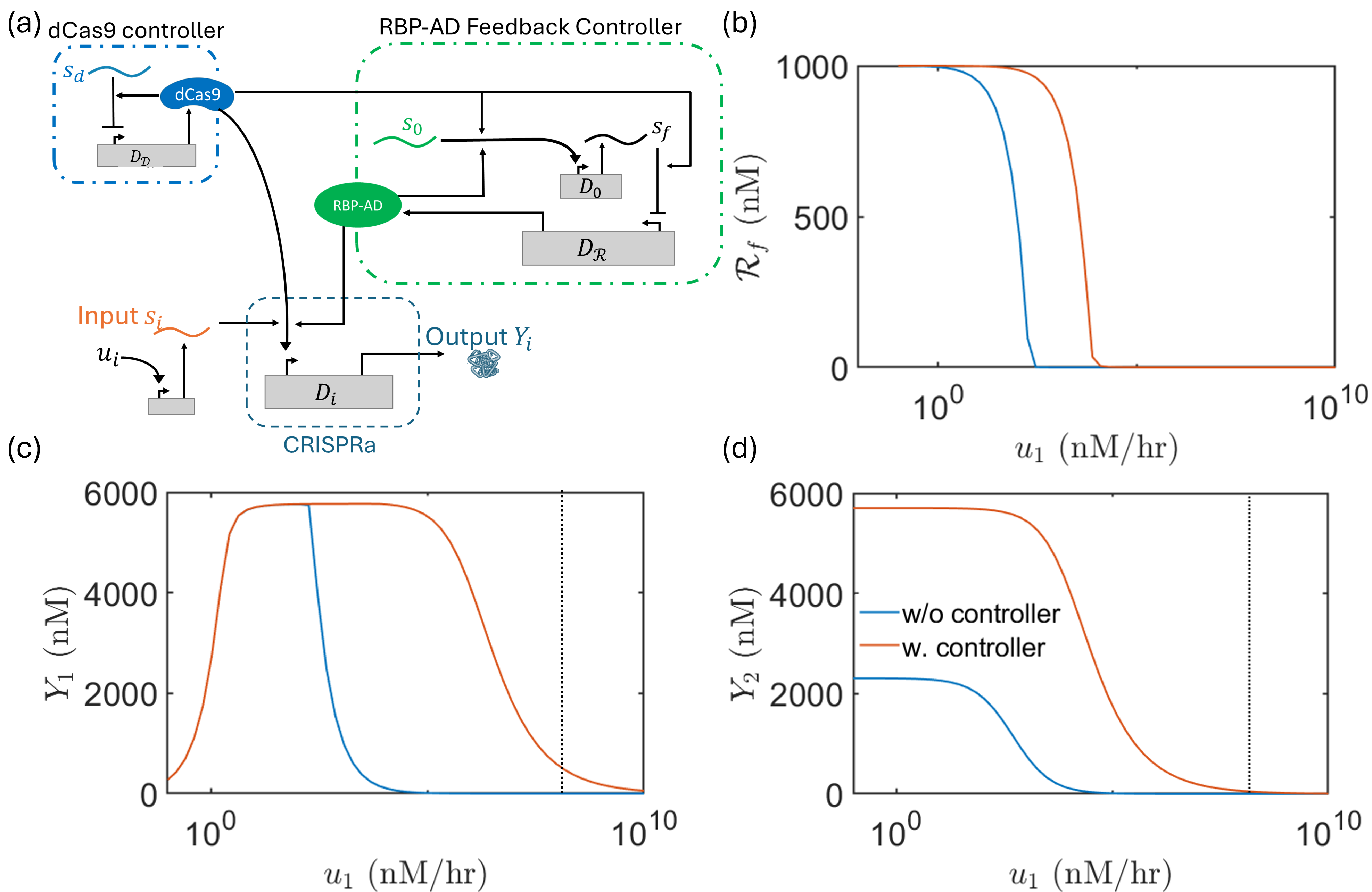}
\caption{Efficacy of the feedback controller for RBP-AD with the dcas9 controller designed in \cite{huang2021dcas9}. (a) Genetic circuit diagram for the CRISPRa module with the feedback controller for RBP-AD (from Section \ref{fcontroller}) and feedback controller for dCas9 (from Section \ref{dcontroller}).The concentration levels of (b) free RBP-AD resource $\r_f$ (c) output protein $Y_1$, (d) output protein $Y_2$, as the input $u_1$ is varied with ($u_0 = 10^8$ nM/hr and $u_d = 10^{10}$ nM/hr) and without ($u_0 = 0$ nM/hr and $u_d = 0$ nM/hr) the controller. $u_2 = 0$ nM/hr for (b) and (c), while $u_2 = 500$ nM/hr for (d). The vertical lines in (c) and (d) correspond to $u_0 = 10^8$ nM/hr.}
\label{fig:feedback:dcas9}
\end{figure}
Contrary to the buffering controller, dCas9 is used in the design of the feedback controller (Section \ref{fcontroller}). Due to the use of dCas9, the transcription rate $\kappa_d$ for dCas9 controller is modified to: 
\begin{align*}
    \kappa_d &= (\d_t + \B{A}_{1,d} + \frac{\B{A}_{1,d} D_{\d t}}{(K_t + \B{A}_{1,d})}) \\
    &\qquad + \B{A}_{1,0} + \B{A}_{1,f} + \B{A}_f + \B{c}_f + \B{C}_f) \frac{\gamma (K_t + \B{A}_{1,d})}{D_{\d t}K_t}.
\end{align*}
This ensures that in the absence of load $u_i = 0$ for all $i \in \{1,n\}$, $\B{\d}_f$ is the same as $\d_t$, for a fair comparison of the CRISPRa system without any controllers and the CRISPRa system with feedback controller for RBP-AD and dCas9 regulation (Fig. \ref{fig:feedback:dcas9}a). The numerical simulations on RBP-AD feedback controller are shown in Fig. \ref{fig:feedback:dcas9}. Fig. \ref{fig:feedback:dcas9} (b) shows a decrease in the RBP-AD levels in the presence of the dCas9 controller when compared to keeping dCas9 constant (see Fig. \ref{fig:feedback}b). Furthermore, a similar adverse shift is observed in the self-sequestration behavior (Fig. \ref{fig:feedback:dcas9}c) and the off-target response (Fig. \ref{fig:feedback:dcas9}d) when compared to the case with the constant dCas9 (Fig. \ref{fig:feedback}c,d).

Comparing the performance of the buffering and the feedback controller on the on-target performance, the value of $u_1$ at which the output $Y_1$ begins to decrease is higher for the buffering controller. Hence, the buffering controller maintains the $Y_1$ value until larger $u_1$ compared to the feedback controller before the self-sequestration sets in (compare Fig. \ref{fig:feedback:dcas9}c with Fig. \ref{fig:buffer:dcas9}c). This is contrary to the case when the dCas9 levels are maintained constant (compare Fig. \ref{fig:feedback}c with Fig. \ref{fig:buffer}c). Moreover, the buffering controller is more efficient at removing off-target effects when compared to a feedback controller (compare Fig. \ref{fig:feedback:dcas9}d with Fig. \ref{fig:buffer:dcas9}d).

\begin{figure}[ht]
\centering
\includegraphics[width=0.47\textwidth]{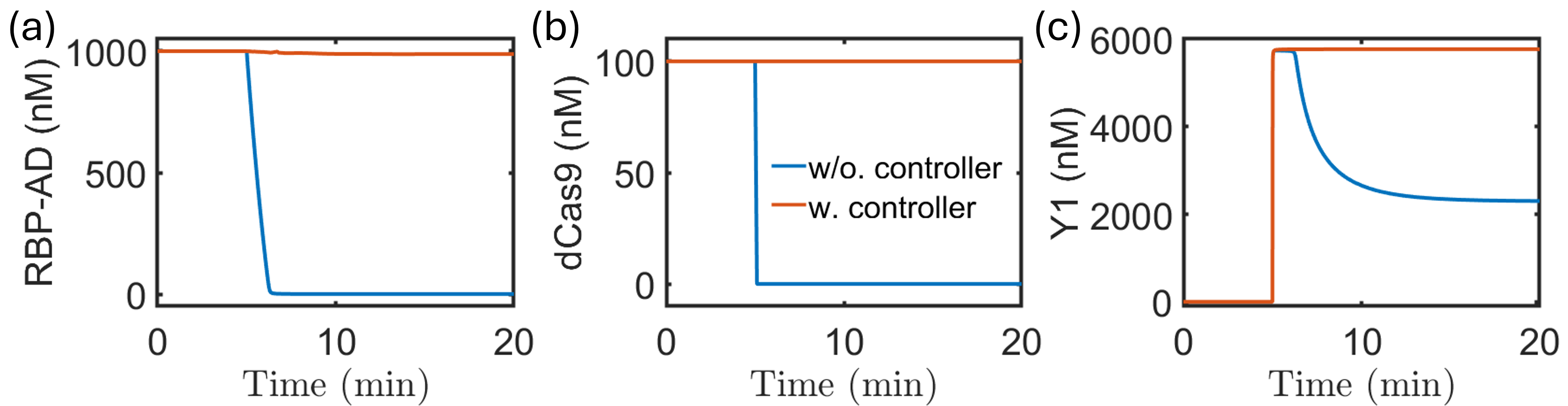}
\caption{Temporal dynamics with and without the controller. Time series data for the concentration levels of (a) RBP-AD ($\r_b$), (b) dCas9 ($\d_f$), (c) output protein $Y_1$, for $u_b = 10^8$ nM/hr, $u_d = 10^{10}$ nM/hr and $n = 1$. The inputs to the CRISPRa modules are $u_1 = 0 \text{ nM/hr}$ for time $t<5$ minutes and $u_1= 100 \text{ nM/hr}$ for time $t\geq 5$ minutes.} \label{fig:timeseries}
\end{figure}

Having established that the buffering controller outperforms the feedback controller when combined with the dCas9 controller, we now examine the temporal response of the system. Figure \ref{fig:timeseries} displays the time course of the controlled resources RBP-AD and dCas9, regulated using buffering and feedback controllers, respectively. Without the controller, the levels of the resources RBP-AD and dCas9 drop when the input $u_1$ is introduced to the system (Fig. \ref{fig:timeseries}a, b). By contrast, their concentrations show negligible change when regulated with the controller. Furthermore, the drop in the output protein concentration is mitigated when the controller is in action (Fig. \ref{fig:timeseries}c).

\section{CONCLUSIONS}\label{sec4}
CRISPRa has been widely used for the parallel activation of multiple genes for genetic screening and therapeutics \cite{kampmann2018crispri,becirovic2022maybe}. Resource competition for dCas9 and RBP-AD, however, causes performance degradation and thus may affect the outcomes in these applications. In this paper, we have designed a multi-variable controller for attenuating competition effects in CRISPRa networks. The controllers were designed based either on a buffering approach or a feedback mechanism. The feedback controller outperforms the buffering one in maintaining the sensitivity of the resources to loads as low.
On the other hand, the concurrent operation of the RBP-AD buffering controller and dCas9 feedback controller shows that the on-target and the off-target effects were mitigated significantly. Furthermore, the orthogonality of the buffering controller with the dCas9 controller facilitates easier experimental implementation and validation. This work then sets the foundations for future two-variable controllers to make CRISPRa networks resilient to loads.

\addtolength{\textheight}{-12cm}   % This command serves to balance the column lengths

%%%%%%%%%%%%%%%%%%%%%%%%%%%%%%%%%%%%%%%%%%%%%%%%%%%%%%%%%%%%%%%%%%%%%%%%%%%%%%%%

%%%%%%%%%%%%%%%%%%%%%%%%%%%%%%%%%%%%%%%%%%%%%%%%%%%%%%%%%%%%%%%%%%%%%%%%%%%%%%%%

%%%%%%%%%%%%%%%%%%%%%%%%%%%%%%%%%%%%%%%%%%%%%%%%%%%%%%%%%%%%%%%%%%%%%%%%%%%%%%%%
%\section*{APPENDIX}

%\section*{ACKNOWLEDGMENT}

%%%%%%%%%%%%%%%%%%%%%%%%%%%%%%%%%%%%%%%%%%%%%%%%%%%%%%%%%%%%%%%%%%%%%%%%%%%%%%%%

\bibliographystyle{IEEEtran}
\bibliography{References}

\begin{thebibliography}{10}
\providecommand{\url}[1]{#1}
\csname url@rmstyle\endcsname
\providecommand{\newblock}{\relax}
\providecommand{\bibinfo}[2]{#2}
\providecommand\BIBentrySTDinterwordspacing{\spaceskip=0pt\relax}
\providecommand\BIBentryALTinterwordstretchfactor{4}
\providecommand\BIBentryALTinterwordspacing{\spaceskip=\fontdimen2\font plus
\BIBentryALTinterwordstretchfactor\fontdimen3\font minus \fontdimen4\font\relax}
\providecommand\BIBforeignlanguage[2]{{%
\expandafter\ifx\csname l@#1\endcsname\relax
\typeout{** WARNING: IEEEtran.bst: No hyphenation pattern has been}%
\typeout{** loaded for the language `#1'. Using the pattern for}%
\typeout{** the default language instead.}%
\else
\language=\csname l@#1\endcsname
\fi
#2}}

\bibitem{mooney2001evolutionary}
H.~A. Mooney and E.~E. Cleland, ``The evolutionary impact of invasive species,'' \emph{Proceedings of the National Academy of Sciences}, vol.~98, no.~10, 2001.

\bibitem{gulley2018china}
A.~L. Gulley, N.~T. Nassar, and S.~Xun, ``China, the united states, and competition for resources that enable emerging technologies,'' \emph{Proceedings of the national academy of sciences}, vol. 115, no.~16, 2018.

\bibitem{durham1976resource}
W.~H. Durham, ``Resource competition and human aggression, part i: a review of primitive war,'' \emph{The Quarterly Review of Biology}, vol.~51, no.~3, 1976.

\bibitem{gyorgy2014limitations}
A.~Gyorgy and D.~Del~Vecchio, ``Limitations and trade-offs in gene expression due to competition for shared cellular resources,'' in \emph{53rd IEEE Conference on Decision and Control}, 2014, pp. 5431--5436.

\bibitem{gyorgy2015isocost}
A.~Gyorgy, J.~I. Jim{\'e}nez, J.~Yazbek, H.-H. Huang, H.~Chung, R.~Weiss, and D.~Del~Vecchio, ``Isocost lines describe the cellular economy of genetic circuits,'' \emph{Biophysical journal}, vol. 109, no.~3, 2015.

\bibitem{al2022epigenetic}
M.~A. Al-Radhawi, S.~Tripathi, Y.~Zhang, E.~D. Sontag, and H.~Levine, ``Epigenetic factor competition reshapes the emt landscape,'' \emph{Proceedings of the National Academy of Sciences}, vol. 119, no.~42, 2022.

\bibitem{Bikard2013}
D.~Bikard, W.~Jiang, P.~Samai, A.~Hochschild, F.~Zhang, and L.~A. Marraffini, ``Programmable repression and activation of bacterial gene expression using an engineered {CRISPR}-cas system,'' \emph{Nucleic Acids Research}, vol.~41, no.~15, June 2013.

\bibitem{larson2013crispr}
M.~H. Larson, L.~A. Gilbert, X.~Wang, W.~A. Lim, J.~S. Weissman, and L.~S. Qi, ``Crispr interference (crispri) for sequence-specific control of gene expression,'' \emph{Nature protocols}, vol.~8, no.~11, 2013.

\bibitem{adli2018crispr}
M.~Adli, ``The crispr tool kit for genome editing and beyond,'' \emph{Nature communications}, vol.~9, no.~1, pp. 1--13, 2018.

\bibitem{Zhang2018}
S.~Zhang and C.~A. Voigt, ``Engineered {dCas}9 with reduced toxicity in bacteria: implications for genetic circuit design,'' \emph{Nucleic Acids Research}, 2018.

\bibitem{chen2018model}
P.-Y. Chen, Y.~Qian, and D.~Del~Vecchio, ``A model for resource competition in crispr-mediated gene repression,'' in \emph{IEEE Conference on Decision and Control (CDC)}, 2018.

\bibitem{anderson2021competitive}
D.~A. Anderson and C.~A. Voigt, ``Competitive dcas9 binding as a mechanism for transcriptional control,'' \emph{Molecular Systems Biology}, vol.~17, no.~11, 2021.

\bibitem{huang2021dcas9}
H.-H. Huang, M.~Bellato, Y.~Qian, P.~C{\'a}rdenas, L.~Pasotti, P.~Magni, and D.~Del~Vecchio, ``dcas9 regulator to neutralize competition in crispri circuits,'' \emph{Nature communications}, vol.~12, no.~1, 2021.

\bibitem{Fontana2018}
J.~Fontana, C.~Dong, J.~Y. Ham, J.~G. Zalatan, and J.~M. Carothers, ``Regulated expression of {sgRNAs} tunes {CRISPRi} in e. coli,'' \emph{Biotechnology Journal}, vol.~13, no.~9, 2018.

\bibitem{manoj2022emergent}
K.~Manoj and D.~Del~Vecchio, ``Emergent interactions due to resource competition in crispr-mediated genetic activation circuits,'' in \emph{IEEE 61st Conference on Decision and Control (CDC)}, 2022.

\bibitem{del2014biomolecular}
D.~Del~Vecchio and R.~M. Murray, \emph{Biomolecular feedback systems}.\hskip 1em plus 0.5em minus 0.4em\relax Princeton University Press, 2014.

\bibitem{cahill1994regulatory}
M.~A. Cahill, W.~H. Ernst, R.~Janknecht, and A.~Nordheim, ``Regulatory squelching,'' \emph{FEBS letters}, vol. 344, no. 2-3, 1994.

\bibitem{hancock2017interplay}
E.~J. Hancock, J.~Ang, A.~Papachristodoulou, and G.-B. Stan, ``The interplay between feedback and buffering in cellular homeostasis,'' \emph{Cell systems}, vol.~5, no.~5, 2017.

\bibitem{kampmann2018crispri}
M.~Kampmann, ``Crispri and crispra screens in mammalian cells for precision biology and medicine,'' \emph{ACS chemical biology}, vol.~13, no.~2, 2018.

\bibitem{becirovic2022maybe}
E.~Becirovic, ``Maybe you can turn me on: Crispra-based strategies for therapeutic applications,'' \emph{Cellular and Molecular Life Sciences}, vol.~79, no.~2, 2022.

\end{thebibliography}

\end{document}